\documentclass[twocolumn,prl,superscriptaddress]{revtex4}

\usepackage[dvips]{graphicx}
\usepackage{amsmath}

\newcommand{\dd}{\mathrm{d}}
\newcommand{\iu}{\mathbf{i}}

\begin{document}

\date{\today}

\title{Isotope effect on superconductivity in Josephson coupled
stripes in underdoped cuprates}

\author{A. Rosengren}%
\email{roseng@kth.se}
\affiliation{Condensed Matter Theory, Department of Theoretical Physics,
AlbaNova University Center, KTH, SE-106 91 Stockholm, Sweden}%
\affiliation{NORDITA, AlbaNova University Center, KTH, SE-106 91 Stockholm,
Sweden}%

\author{P. H. Lundow}%
\email{phl@kth.se}
\affiliation{Condensed Matter Theory, Department of Theoretical Physics,
AlbaNova University Center, KTH, SE-106 91 Stockholm, Sweden}%

\author{A. V. Balatsky}%
\email{avb@lanl.gov} \affiliation{Theoretical Division, MS B262, Los
Alamos National
Laboratory, Los Alamos, New Mexico 87545, USA}%

\begin{abstract}
  Inelastic neutron scattering data for YBaCuO as well as for LaSrCuO
  indicate incommensurate neutron scattering peaks with
  incommensuration $\delta(x)$ away from the $(\pi,\pi)$ point.
  $T_c(x)$ can be replotted as a linear function of the
  incommensuration for these materials. This linear relation implies
  that the constant that relates these two quantities, one being the
  incommensuration (momentum) and another being $T_c(x)$ (energy), has
  the dimension of velocity we denote $v^*$: $k_B T_c(x) = \hbar v^*
  \delta(x)$. We argue that this experimentally derived relation can
  be obtained in a simple model of Josephson coupled stripes. Within
  this framework we address the role of the $O^{16} \rightarrow
  O^{18}$ isotope effect on the $T_c(x)$. We assume that the
  incommensuration is set by the {\em doping} of the sample and is not
  sensitive to the oxygen isotope given the fixed doping. We find
  therefore that the only parameter that can change with O isotope
  substitution in the relation $T_c(x) \sim \delta(x)$ is the velocity
  $v^*$.  We predict an oxygen isotope effect on $v^*$ and expect it
  to be $\simeq 5\%$.

\end{abstract}

\maketitle

\section{Introduction}
The isotope effect has played an important role in the understanding
of the underlying pairing mechanism in superconductors.  Historically
it was used to identify the role of the electron-lattice interaction
for the superconductivity. Experimental evidence as to the nature of
interaction that causes superconductivity were first provided in 1950
by Maxwell~\cite{maxwell:50} and by Reynolds et
al.~\cite{reynolds:50}.  They showed that $T_c \propto M^{-\alpha}$,
where $\alpha = 0.5 \pm 0.05$ and $M$ is the mean mass of different
isotopes of the superconductor.  These findings indicated that the ion
mass, and therefore lattice vibrations, phonons, are important to the
mechanism of superconductivity.

Fr\"ohlich~\cite{frohlich:50}, who was unaware of the experiments on
the isotope effect, and Bardeen~\cite{bardeen:50}, the same year,
have provided theories of the phonon-electron interaction, which in
turn led to models of superconductivity dependent on the phonon
energies.

For the high-$T_c$ superconductors the study of the isotope effect
does not paint a simple and straightforward picture.  The role of the
electron-lattice interactions in the mechanism of superconductivity
was initially ruled out, and the pairing mechanism was ascribed to
antiferromagnetic exchange and
fluctuations~\cite{anderson:87,scalapino:95}. As time goes by we
witness the growing acknowledgement that interactions of the lattice
with the carriers in high-$T_c$ might be important. In the discussion,
the role of lattices and phonons is today gaining importance
again~\cite{gweon:04, bussmann:05, lee:06, bishop:07, hwang:07,
douglas:07, gweon:07,wang:07}.  The isotope effect on $T_c$ is
believed to be small at optimal doping, but increases to the BCS value
in the underdoped regime~\cite{franck:94}. What complicates the
discussion on the isotope effect is the fact that underdoped
LSCO~\cite{valla:06} is electronically inhomogeneous. Inhomogeneity is
also well established in a Bi2212 superconductor~\cite{lee:06}.  The
situation is different for the YBCO compounds that are believed to be
more homogeneous. In both the LSCO and the YBCO case the
incommensuration, that possibly is related to stripes, is certain.

For the purposes of this discussion, we would like to point out a
distinction of the isotope effect we consider here versus the notion
of an isotope effect in the inhomogeneous systems. In the case of
conventional homogeneous superconductors, a discussion of the isotope
effect is centered on an exponent that describes the effect of ionic
isotope substitution on the superconducting $T_c$. In the case of
spatially inhomogeneous systems and materials with more than one
energy scale, e.g. superconducting gap vs pseudogap energy scale, the
notion of an isotope effect has to be expanded to address the
difference in changes that could be caused by isotope substitution on
different energy scales~\cite{lanzara:99,rubio:02, bussmann:05b,
keller:05, furrer:05, sury:05,hafliger:06, muller:07}.
Similar arguments can be made about the effect isotope substitution
can have on pairing gap vs the superfluid stiffness. The very notion
of a {\em single exponent} for the isotope effect in the presence of
an electronic inhomogeneity that characterizes the whole sample by a
single exponent, has to be viewed at best as a very crude average
description of what is really happening in these materials.

We will take the view that there are stripes in the underdoped
cuprates and address how they modify the isotope effect.  The
discussion on the precise real space shape of the stripes in the
presence of disorder has revealed a variety of complicated patterns
\cite{kivelson:03, robertson:06}. We are not concerned here with the
specific form of stripe order since we assume some typical
stripe-stripe distance.

Recent scanning tunneling microscopy data on two lightly hole-doped
cuprates, Ca$_{1.88}$Na$_{0.12}$CuO$_2$Cl$_2$ and
Bi$_2$Sr$_2$Dy$_{0.2}$Ca$_{0.8}$Cu$_2$O$_{8+x}$ by Kohsaka et
al.~\cite{kohsaka:07}, reported the presence of a cluster glass with a
large pairing amplitude of the localized pairs on the oxygen sites
that form a real space glass-like pattern.  We therefore assume that
superconductivity in the underdoped regime develops through the onset
of phase coherence between superconducting regions that communicate
with each other via Josephson coupling \cite{eroles:00}.  The precise
real space arrangement of these regions is not crucial for our
analysis except for the fact that the SC regions look like quasi-1D
clusters with broken orientational symmetry.

In this paper we discuss the effect of isotope substitution on
Josephson coupled stripes. In doing so our starting point will be the
linear relation between the incommensurate peak splitting and $T_c$
observed in YBCO and LSCO \cite{yamada:98,dai:01}. In case of YBCO
this relation does not extend as far as for LSCO as a function of
doping.

Inelastic neutron scattering data for YBCO as well as for LSCO
indicate incommensurate neutron scattering peaks with incommensuration
$\delta(x)$ away from the $(\pi,\pi)$ point.  It is also known that
$T_c(x)$ taken as a function of doping $x$ can be replotted as a
linear function of the incommensuration for these materials, a so
called Yamada plot. This proportionality implies that the constant
that relates these two quantities, one being the incommensuration
(momentum) and another being $T_c(x)$, or energy, has the dimension of
velocity and is denoted $v^*$:
\begin{equation}\label{eq0}
  k_B\, T_c(x) = \hbar\,v^*\,\delta(x),
\end{equation}
This experimentally derived relation can be obtained in a simple model
of Josephson coupled stripes.  We address the role of the $O^{16}
\rightarrow O^{18}$ isotope effect on the $T_c(x)$ within this
framework. We argue that the incommensuration is set by the
\emph{doping} of the sample and is not sensitive to the oxygen isotope
given the fixed doping. We find therefore that the only parameter that
can change in the relation $T_c(x) \sim \delta(x)$ is the velocity
$v^*$.  We estimate that the effect of isotope substitution on $v^*$
is on the order of $5\%$ for both LSCO and YBCO materials.


\section{Discussion}
Progress in neutron scattering has allowed for a multitude of
inelastic neutron scattering data to be gathered for the high-$T_c$
superconductor YBa$_2$Cu$_3$O$_{6+x}$.  If one follows the
off-resonance spectrum to lower energies one finds incommensurate
peaks with an incommensuration $\delta$ that is doping dependent.
>From the neutron data for YBCO for oxygen concentration $x$, where
$0.45\le x \le 0.95$, with $\max T_c(x) = 93\, \textrm{K}$, a simple
linear relation between $T_c$ and $\delta $ for the doping range $x
\le 0.6$ was found to follow Eq.~\ref{eq0} with $\hbar\,v^* =
37\,\textrm{meV {\AA}}$, see~\cite{dai:01}.


Another well-studied system is LaSrCuO. Inelastic neutron scattering
on La$_{214}$ compounds show incommensurate peaks at $(\pi\pm
\delta,\pi)$ and $(\pi,\pi\pm\delta)$, see~\cite{yamada:98}.  It was
found that $T_c$ was a linear function of $\delta$ up to the optimal
Sr doping value~\cite{yamada:98} and Eq.~\ref{eq0} holds as well. Here
the constant of proportionality for LSCO is $\hbar\,v^* =
20\,\textrm{meV \AA}$. For both materials the velocity $\hbar\,v^*$ is
two orders of magnitude smaller than the Fermi velocity of nodal
quasiparticles $\hbar\,v_F \simeq 1\,\textrm{eV \AA}$,
see~\cite{balatsky:99}. Further the velocity $\hbar\,v^*$ is one order
of magnitude smaller than the spin-wave velocity
$\hbar\,v_{\textrm{sw}} \simeq 0.65\,\textrm{eV \AA}$ of the parent
compound~\cite{yamada:98}. Similarly, for La$_{214}$ the inferred
velocity is much smaller than the spin-wave velocity
$\hbar\,v_{\textrm{sw}} \simeq 0.85\,\textrm{eV \AA}$,
see~\cite{aeppli:89}.

\begin{figure}
  \begin{center}
    \includegraphics[width=0.48\textwidth]{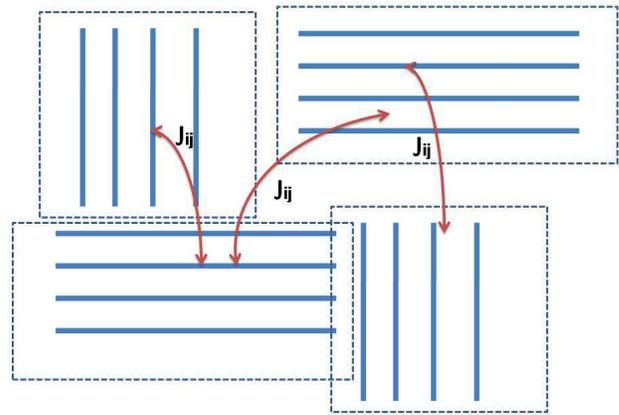}
  \end{center}
  \caption{Josephson coupling $J_{ij}$ between grains that contain
    stripe pieces is shown schematically. The average J determined by
    averaging over an ensemble of $J_{ij}$ will determine $T_c$.}
  \label{FigJ}
\end{figure}

\begin{figure}
  \begin{center}
    \includegraphics[width=0.48\textwidth]{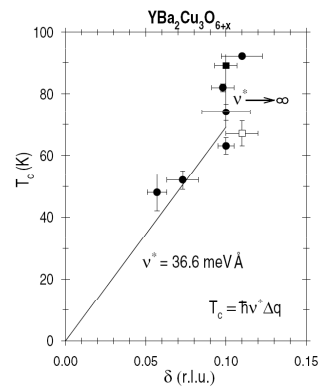}
  \end{center}
  \caption{Linear relation between the incommensurate peak splitting
  $\delta$ and $T_c$ for the YBCO superconductor. The linear slope of
  the curve means that $k_B\, T_c(x) = \hbar\,v^*\,\delta$. We find
  $\hbar\,v^* = 37\,\textrm{meV {\AA}}$. The figure is taken from
  Fig. 25 of \cite{dai:01}.}
  \label{Figslope}
\end{figure}

Does the relation in Eq.~\ref{eq0} imply the existence of an
excitation with such a velocity? An interpretation of this relation is
to connect the superconductivity mechanism to the existence of
fluctuating stripes.  The simple relation above gives an inverse
proportionality between $T_c(x)$ and the doping dependent length
$\ell(x)$ determined from neutron scattering, $\ell(x) = 1/\delta(x)$.
Josephson tunneling of pairs between stripe segments can produce such
a relation.

A model Hamiltonian of random stripe separation and inter- and
intra-stripe random Josephson coupling is
\begin{equation}\label{eq3}
  \mathcal{H} = \sum_{ij} J_{ij} \, \exp\left(\iu\left(\phi_i -
  \phi_j\right)\right),
\end{equation}
where the summation is taken over coarse-grained regions $i$ with
well-defined phases, and where $J_{ij} = J(r_{ij}) = t_0 /
r_{ij}^{\beta}$.  It is here assumed that $J(r)$ has an exponential
cutoff at lengths much larger than the stripe-stripe distance in order
to have a well-defined thermodynamic limit.  The stripe-stripe
distance $r$ is given by a probability distribution $P(r,\theta)$.
For simplicity we will, as was done in \cite{eroles:00}, assume a
uniform distribution in 2D with $P(r,\theta) = C$ for $\ell-a \le r
\le \ell+a$ and $0\le\theta<2\pi$, otherwise $P(r,\theta) = 0$.  Here
$a=\nu\ell$ where $\nu$ is a bounded parameter. This gives

\begin{gather*}
  \int_0^{2\pi}\!\!\int_0^{\infty}\! P(r,\theta)\, \dd r\, \dd\theta
  =2\,\pi\,C\, \int_{\ell-a}^{\ell+a}\! r\,\dd r = \\
  = 4\,\pi\,C\,\ell\, a = 1
\end{gather*}
so that $C = (4\,\pi\,\ell\,a)^{-1}$. This gives the expected $\langle
r \rangle$
\begin{gather*}
  \langle r \rangle = \int_0^{2\pi}\!\!\int_0^{\infty}\! r\,P(r,\theta)
  \dd r\,\dd\theta =
  2\,\pi\,C\,\int_{\ell-a}^{\ell+a}\! r^2\,\dd r = \\
  = \ell + \frac{a^2}{3\,\ell} \propto \ell
\end{gather*}

and the expected $\langle J(r) \rangle$
\begin{gather*}
  \langle J(r) \rangle =
  \langle t_0 / r^{\beta} \rangle =
  \int_0^{2\pi}\!\!\int_0^{\infty}\! t_0\,r^{-\beta}\,P(r,\theta)\,
  \dd r\,\dd\theta = \\
   = 2\,\pi\,C\,t_0\,\int_{\ell-a}^{\ell+a}\! r^{1-\beta}\,\dd r =\\
  \frac{2\,\pi\,C\,t_0}{2-\beta}\left((\ell+a)^{2-\beta} -
  (\ell-a)^{2-\beta}\right)
\end{gather*}
which for $\beta=1$ gives $\langle J(r) \rangle = t_0/\ell$ so that
one recovers the experimentally observed relation~\cite{eroles:00}
\begin{equation}\label{eq4}
  T_c(x) \simeq \langle J(r)\rangle \propto \langle r \rangle^{-1}
  \propto\delta(x).
\end{equation}
The velocity $v^*$ cannot be determined for this simple model without
any further assumptions. We suggest that $v^*$ is related to the phase
dynamics of the superconducting regions (stripes).

>From the simple relation in Eq.~\ref{eq0} we can now investigate the
effect of isotope substitution on $T_c$. Because the hole
concentration is not changing and since $\delta(x)$ is not changing
with isotope substitution, the only parameter left to be isotope
dependent is $\hbar\,v^*$. Since $v^*$ is related to the phase
dynamics of the stripes it is natural to expect that $v^*$ will not
change much by isotope substitution, because of its slight effect on
the band structure. From the measured oxygen isotope effect on $T_c$
of YBCO~\cite{bornemann:91}, we predict
$v_{18}^* / v_{16}^*$,to be at least $0.95$,
where the velocity $v^*$ has been indexed by the isotope mass, in
agreement with our expectation.

The prediction on the change of $v^*$ with isotope is made as
follows. The isotope effect parameter $\alpha$ is calculated as
\[
\alpha = \frac{
  \ln\left(1-(T_c^{16}-T_c^{18})/T_c^{16}\right)
}
       {
     \ln\left(m_{16}/m_{18}\right)
       },
       \]
where $m_{16}$ and $m_{18}$ are the oxygen isotope
masses. Further, due to Eq.~\ref{eq0},  $T_c^{18}/T_c^{16} =
v_{18}^* / v_{16}^*$. This leads to
\[
\frac{v_{18}^*}{v_{16}^*}=\left(\frac{16}{18}\right)^{\alpha}.
\]
For YBCO $\alpha = 0.27$ ($T_c^{16} = 60\,
\textrm{K}$)~\cite{bornemann:91} so we get $v_{18}^* / v_{16}^* =
0.969$, for LSCO $\alpha = 0.38$ ($T_c^{16} = 38.3\, \textrm{K}
$)~\cite{bornemann:91} and so $v_{18}^* / v_{16}^* = 0.956$.

We find that if the doping level is kept the same in isotope
substitution then the typical stripe-stripe distance, controlled by
doping $x$, does not change with $x$. Therefore the only parameter
that can change with isotope substitution is the coefficient that
relates $T_c$ to $\delta(x)$: $k_B T_c(x) = \hbar v^* \delta(x)$.
This coefficient $v^*$ has dimension velocity and describes the phase
dynamics in the Josephson coupled superconductors. Since $v^*$ is
related to an electronic degree of freedom, it is hardly surprising
that it is only weakly dependent on O isotope substitution.  We
estimated the isotope effect on $v^*$, or equivalently on $T_c$, to be
less than $5\%$. We can therefore predict the change of $v^*$ to be of
the order of a few percent in a wide doping range. This estimate is
consistent with the isotope effect observed for the superfluid
stiffness $\rho_s$ for underdoped LCSO~\cite{khasanov:04,yusupov:06}.

\section{Conclusion}

In conclusion, we have considered the role of the $O^{16} \rightarrow
O^{18}$ isotope effect on the $T_c$ of the Josephson coupled stripes
and its implications for $v^*$. We find that the effect is small and
is on the order of $5\%$ at most. We argue that the effect on $v^*$ is
small because the underdoped materials enter into a superconducting
state due to phase fluctuations and therefore the main effect that
controls $T_c$ is a Josephson coupling between superconducting
regions. If these phase fluctuations are due to electronic degrees of
freedom, lattice dynamics has a small but observable effect on $v^*$
that we estimate to be on the order of $\simeq 5\%$. This estimate
would imply a similar scale effect on superfluid stiffness $\rho_s$,
an estimate that is consistent with other experiments
\cite{khasanov:04}.

\section{Acknowledgements}

We are grateful to J.C. Davis, T. Deveraux, Y. Koshsaka, Z.X. Shen,
J.X. Zhu for useful discussions.  This work was supported by US DOE
BES, and the Swedish Research Council. Two of the authors (AR, PHL)
thank the Theoretical Division, Los Alamos National Laboratory for the
kind hospitality.

\bibliography{isotopes-final[2]}

\end{document}